# SLOWDOWN OF MICROPARTICLES BY AN ELECTROMAGNETIC POTENTIAL WELL DEEPENING OVER TIME

Azad Ch. Izmailov

Institute of Physics, Azerbaijan National Academy of Sciences, Javid av. 33, Baku, Az-1143, AZERBAIJAN

*e-mail*: azizm57@rambler.ru

***ABSTRACT***. We analyze possible motion control of microparticles by means of external electromagnetic fields which induce (for such particles) potential wells having fixed spatial distribution but deepening over time up to some limit. It is assumed that given particles are under conditions of the high vacuum and forces acting on these particles are not dissipative. We have established slowdown of comparatively fast particles as a result of their transit through considered potential wells. This process is demonstrated on example of the nonresonance laser beam with the intensity amplifying over time. More detailed research of particle slowdown in such electromagnetic fields is carried out on the basis of simple analytical relationships obtained from basic equations of classical mechanics for the model of the one-dimensional rectangular potential well deepening over time. Method for "cooling" of particles, demonstrated in the present work, may be applied for essential increase of spectroscopy resolution of various microparticles, including in definite cases also atoms and molecules in the ground quantum state.

# 1. INTRODUCTION

Effective motion control of various microparticles (in particular atoms, molecules and ions) at conditions of high vacuum by external electromagnetic fields is very important for different directions of physics and technique including, of course, spectroscopy of such particles [1-3].

In papers [4,5] we have established the new trapping mechanism of sufficiently slow classical particles in electromagnetic potential wells, which have fixed spatial distribution but deepen over time (up to some limit). Situations were considered when given particles were under conditions of the high vacuum and forces acting on given particles were not dissipative (that is motion of these particles occurred without friction). Depending on whether particles have electric or magnetic moment, it is possible to use the controllable electric or magnetic field and also nonresonance laser radiation for their trapping by such a method. Corresponding electromagnetic traps of particles without material walls make localization and observation of these particles possible for a relatively long time, thus creating conditions for detailed analysis of their properties. According to paper [4], such a trapping at definite conditions may be realized also for atoms and molecules in the ground quantum state.

In the present work we show possible slowdown of comparatively fast particles as a result of their transit through similar electromagnetic potential wells (section 2). Such a slowdown is demonstrated on example of the nonresonance laser beam with the intensity amplifying over time (section 3). In general case, dynamics of particles in considered fields may be calculated only numerically. At the same time, much more detailed information on features of particles dynamics may be obtained at analytical solution of corresponding tasks. Such an analysis of particles slowdown is carried out in the present work, where we received simple solution of motion equations of classical particles passing through the one-dimensional rectangular potential well deepening over time (section 4). In conclusion we discuss possible generalization of obtained results and perspectives of their using for high-resolution spectroscopy of various particles including in definite cases also atoms and molecules (section 5).

# 2. BASIC RELATIONSHIPS

As in previous papers [4,5], we will consider problems which may be solved on the basis of classical mechanics and electrodynamics. Let us assume that a point particle with the mass *m* freely moving in a three-dimensional space before its entering to the region *V* of the potential well $U(\boldsymbol{R}, t)$, which explicitly depends not only on the coordinate $\boldsymbol{R}$ but also on time $t$. The total energy of such a particle with the non-relativistic velocity $\boldsymbol{v}$ is described by the known formula [6]:

$$E(\boldsymbol{R}, \boldsymbol{v}, t) = 0.5 m v^2 + U(\boldsymbol{R}, t). \tag{1}$$

Further we will consider the potential energy $U(\boldsymbol{R}, t)$ of the following type:

$$U(\boldsymbol{R}, t) = s(\boldsymbol{R}) * \varphi(t), \tag{2}$$

where the coordinate function $s(\boldsymbol{R}) \leq 0$ in the region $V$, and $\varphi(t) \geq 0$ is the nondecreasing function of time $t$. Such a potential (2) may be created for particles having electric or magnetic moment by a controllable electromagnetic field with the growing strength (up to a certain time moment) but with a fixed spatial distribution [7]. We have the following motion equation of the particle in case of the potential energy (2):

$$m \frac{d^2 \boldsymbol{R}}{dt^2} = -\varphi(t) \frac{ds(\boldsymbol{R})}{d\boldsymbol{R}}. \tag{3}$$

From relations (1)-(3) we directly receive the formula for the time derivative of the total energy $E(\boldsymbol{R}, \boldsymbol{v}, t)$ of the particle:

$$\frac{dE}{dt} = s(\boldsymbol{R}) \frac{d\varphi(t)}{dt} \leq 0. \tag{4}$$

According to inequality (4), increase of the function $\varphi(t)$ with time $t$ leads to decrease of the total energy $E(\boldsymbol{R}, \boldsymbol{v}, t)$ (1) of the particle in the region $V$ of the potential well, where the coordinate function $s(\boldsymbol{R}) \leq 0$. In papers [4,5] we considered the case, when the total energy $E$ of a particle became negative because of the inequality (4). Then, according to the formula (1), such a particle can not overcome the potential well and reach the region with $U(\boldsymbol{R}, t) = 0$. Thus, on the basis of numerical calculations in paper [4], we demonstrated the trapping and localization of classical particles on example of amplifying over time nonresonance standing light wave, whose intensity had the transversal Gaussian distribution.

At the same time, a sufficiently fast particle overcomes such potential wells in spite of their deepening over time. However, according to formulas (1) and (4), kinetic energy of this particle decreases at transits of such wells between space regions where the particle potential energy $U(\mathbf{r}, t) = 0$.

More detailed analysis of particle dynamics may be carried out on the basis of relationships (1)-(4) for electromagnetic potential wells (2) of definite spatial configurations.

## 3. SLOWDOWN OF PARTICLES BY RADIATION OF INCREASING INTENSITY

In this section we will analyze slowdown of considered classical particles by a spatially inhomogeneous electromagnetic radiation with intensity increasing over time up to a certain limit. In so doing, it is assumed that the radiation pressure exerted on particles by such a radiation is small compared to the light induced gradient force acting upon them. This is possible for particles that are nearly transparent in the spectral range of their irradiation. We will analyze not too strong radiation, for which the induced electric dipole moment of the particle is proportional to the electric field strength, while the potential energy of the particle is proportional to the electric field squared [2,7].

Let us analyze the case suitable for practical implementation of a running (along the axis $z$) light beam, the intensity of which increases over time and is characterized by the Gaussian transverse intensity distribution. For particles with induced dipole moment, this radiation creates a potential well of the type (2) with the coordinate function $s(\boldsymbol{R})$ of the form [2,3]:

$$s(\boldsymbol{R}) = -P_0 * exp(-r^2/r_0^2), \qquad (5)$$

where $r = \sqrt{x^2 + y^2}$ is the distance from the beam central axis ($r_0$ is the characteristic beam radius), $P_0 > 0$ is a constant quantity with the dimension of energy that is determined by a particle polarizability. For example, let us consider the following time dependence $\varphi(t)$ (2) for the beam intensity:

$$\varphi(t) = 1 - exp\left(\frac{-t}{\tau}\right), \qquad (t \geq 0), \qquad (6)$$

where $\tau$ is a characteristic time interval. Function $\varphi(t)$ (6) increases with time from 0 to 1 and asymptotically approaches to 1 when $t \gg \tau$.

Fig.1a demonstrates numerically calculated, on the basis of motion equations (3), dependence of the modulus of the particle velocity $v$ on time $t$ at initial conditions specified in a moment $t_0$. The situation is considered, when such particles in this moment $t_0$ are located outside of the region where the light beam (5) can exert sufficient influence on them. In case $t_0 = 0$, particles with the speed $v_0 = v(t_0)$ transit through the laser beam (5) during increasing of its intensity according to the time dependence $\varphi(t)$ (6). Then, at first the sharp increase of the particle speed takes place because of its getting into the light induced potential well (curve 1 in Fig.1a). However, after the passage of such a deepening over time well, the particle speed goes to the constant final value $v_f$, which is noticeably less of its initial value $v_0$. According to the formula (4), this process is accompanied by decrease of the particle total energy (curve 1 in Fig.1b).

Curves 2 in Fig.1 correspond to a particle with the same initial values of velocity and coordinates as for considered above curves 1, but specified in a time moment $t_0 \gg \tau$. Then, according to the time dependence (6), such a particle initially transits through the stationary light beam whose intensity already reaches a maximum value. In this case the final particle speed $v_f$ is equal to its initial value $v_0$ and the total energy of the particle is constant during all process of particle transit through such a stationary potential well (curves 2 in Figs.1a and 1b).

Results of this section were based on numerical solutions of motion equations (1)-(4) for classical particles. At the same time, analytical solutions of given equations for the visual model of the one-dimensional rectangular potential well in the next section 4 allow to reveal features of particles slowdown in considered electromagnetic fields in more details.

## 4. ONE- DIMENSIONAL RECTANGULAR POTENTIAL WELL

Let us consider a point particle with the mass $m$ freely moving with the velocity $v_0 > 0$ along the axis $x$ (Fig.2) from the region $x < -L$ and in a certain moment $t$ reaching the boundary $x = -L$ of the following potential well of the type (2):

$$U(x,t) = -J_0 * \eta(L^2 - x^2) * \varphi(t), \qquad (7)$$

where $J_0 > 0$ is a constant value with the energy dimension, $1 \geq \varphi(t) \geq 0$ is the nondecreasing function of time $t$, $\eta(y)$ is the step function ($\eta(y) = 1$ for $y \geq 0$ and $\eta(y) = 0$ if $y < 0$). From Eq. (3) we receive the motion equation of the particle for the potential well (7):

$$m\frac{d^2x}{dt^2} = J_0 * \varphi(t) * [\delta(x + L) - \delta(x - L)], \qquad (8)$$

where $\delta(y)$ is the Dirac delta-function. According to Eq.(8), an abrupt increase of the particle velocity occurs from the initial value $v_0$ to $v \geq v_0$, when this particle falls into the well (7) in the moment $t$. Connection between given values $v_0$ and $v$ is determined from the formula (1) for the total energy of the classical particle in this moment $t$:

$$E(-L,t) = 0.5mv_0^2 = 0.5mv^2 - J_0 * \varphi(t). \qquad (9)$$

According to Eq.(8), the considered particle further will move inside the rectangular well (7) with the constant velocity $v$. In paper [5] we investigated trapping of particles with sufficiently small initial speed $v_0$ by such a rectangular well (Fig.2). Now we consider the case of comparatively fast particles which overcome this well in spite of its deepening over time. Then we receive from formula (1) the following relationship for the total energy of a particle which reaches the well boundary $x = L$ (Fig.2) in the moment $(t + 2L/v)$:

$$E(L, t + 2L/v) = 0.5mv_f^2 = 0.5mv^2 - J_0 * \varphi(t + 2L/v) > 0, \qquad (10)$$

where $v_f$ is the particle speed after its exit from the well. According to Eq. (10) this final speed $v_f$ has the form:

$$v_f = \sqrt{v^2 - \frac{2J_0}{m} \varphi(t + 2L/v)}, \qquad (11)$$

and the particle speed $v$ in the well is directly determines from Eq. (9):

$$v = \sqrt{v_0^2 + \frac{2J_0 \varphi(t)}{m}}. \qquad (12)$$

Further, for definiteness, we consider the following time dependence $\varphi(t)$ of the potential (7):

$$\varphi(t) = \left(\frac{t}{T}\right)^n \eta(T - t) + \eta(t - T), \quad (n > 0, \ t \geq 0). \qquad (13)$$

According to formula (13), the depth of the potential well (7) increases from 0 to $J_0$ during the period $0 \leq t \leq T$, and will have the maximum constant value $J_0$ when $t > T$. As in paper [5], we introduce following characteristic values for the potential well (7) with dimensions of speed and energy:

$$w = 2L/T, \quad K = 0.5mw^2. \qquad (14)$$

Figures 3 and 4 present dependences of the particle final speed $v_f(t)$ on its initial velocity $v_0$ and on the moment $t$ of particle getting into the well (Fig.2), calculated on the basis of formulas (11) and (12) for the time function $\varphi(t)$ (13). We see that the speed $v_f$ asymptotically approaches to $v_0$ at increase of the value $v_0$ (Fig.3). Indeed a particle with more initial velocity $v_0$ faster overcomes the potential well and is less exposed to its deepening over time. So far as sufficiently slow particles are captured by this well [5], then the final speed $v_f$ of particles passing through the well in Fig.3 takes place only starting from a critical value of the initial speed $v_0$ of a particle.

Fig.4 demonstrates slowdown of particles, determined by the speed difference $(v_0 - v_f)$, on the moment $t$ of particle getting into the well (Fig.2). We see that such a slowdown process will stop and $v_f = v_0$ when time $t > T$. Indeed then, according to relationships (7) and (13), the considered well (Fig.2) will be stationary. It is important to note that the final speed $v_f$ of a particle essentially depends on the parameter $n$ of the function $\varphi(t)$ (13), that is on increase rate of the well depth with time $t$ (Figs.3 and 4).

## 5. CONCLUSION

While continuing our previous papers [4,5], we have investigated dynamics of particles passing through electromagnetic potential wells of the type (2), which deepened with time at their fixed spatial distribution. In this new work we have established and analyzed the slowdown of comparatively fast particles as a result of their transit through such wells. We demonstrated the slowdown of particles on example of the light beam, whose intensity increased with time. Thus it is possible to realize "cooling" of a particle flow crossing nonresonance laser beams during rise of their power. More detailed analysis of particle slowdown we have carried out for the visual model of the rectangular potential well (Fig.2), for which simple analytical relationships of corresponding classical motion equations of particles have been derived. In practice such a rectangular well may be created, for example, by the controllable local homogeneous electric (or magnetic) field on the propagation path of a collimated beam of classical particles having electric (or magnetic) moment.

During a definite time interval of growth of the given field strength (up to a certain maximum value), this electromagnetic well will capture or slow down particles from their beam. Simple analytical relationships obtained in paper [5] and in the present work for the rectangular potential well (Fig.2) may be used for corresponding estimations in studies of dynamics of particles passing also through another electromagnetic wells of the type (2) provided correct definition of an effective width and increasing with time depth of such a well.

We have considered structureless classical particles in potential wells of the certain type (2). In practice it is possible, for example, for a collection of noninteracting (with each other) microparticles, which fly without friction under conditions of the ultrahigh vacuum at action of the controllable electric (magnetic) fields or nonresonance laser radiation with fixed spatial configurations.

For analysis of possible slowdown (cooling) of atoms and molecules at their transit through such electromagnetic wells (2), consideration of their quantum structure is necessary. At the same time, in definite cases results obtained in the present work may be applied also for given atomic objects in the ground quantum state [4].

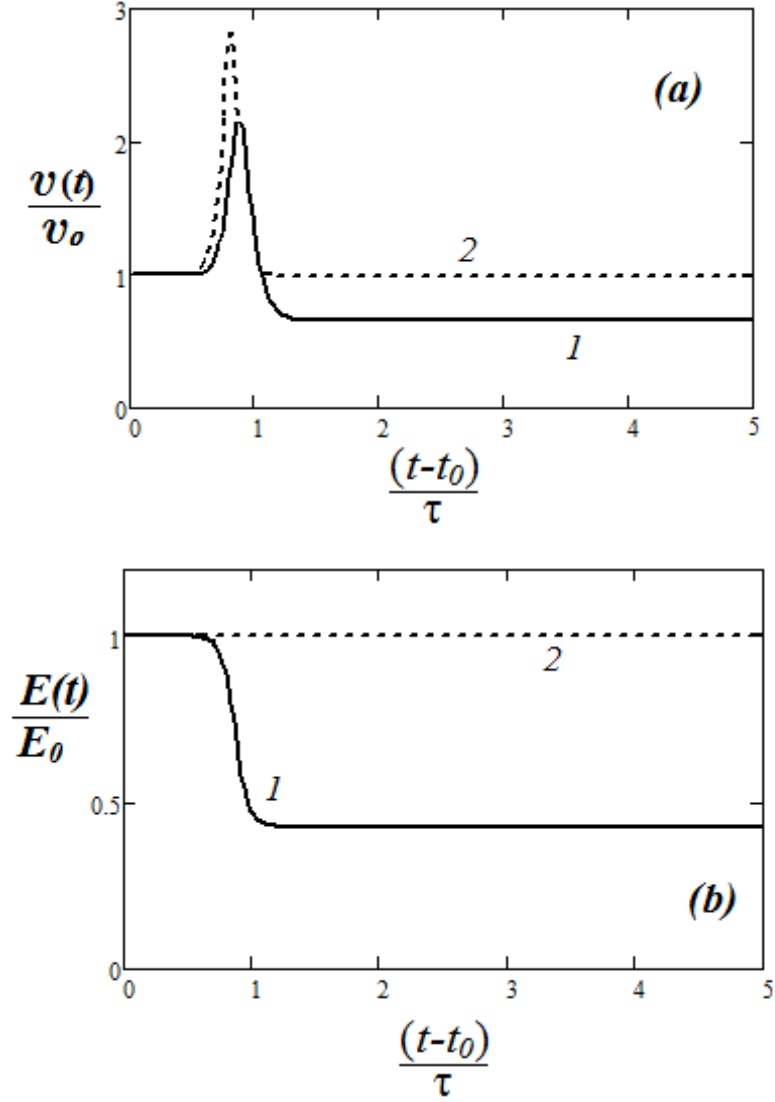

**Fig.1.** Dependence of the particle velocity modulus $v$ (*a*) and its total energy $E$ (*b*) on time $t$ at following initial coordinates of this particle $x(t_0)=0$, $y(t_0) = 5r_0$, $z(t_0)=0$ and its velocity components $v_x(t_0) = 0.5\,(r_0/\tau)$, $v_y(t_0) = -5(r_0/\tau)$ and $v_z(t_0) = 1.5(r_0/\tau)$, specified in the moment $t_0 = 0$ (curves 1) and $t_0 \gg \tau$ (curves 2) for the function $\varphi(t)$ (6), when $P_0 = 100m(r_0/\tau)^2$, $v_0 = v(t_0) = \sqrt{v_x(t_0)^2 + v_y(t_0)^2 + v_z(t_0)^2}$ and $E_0 = E(t_0)$.

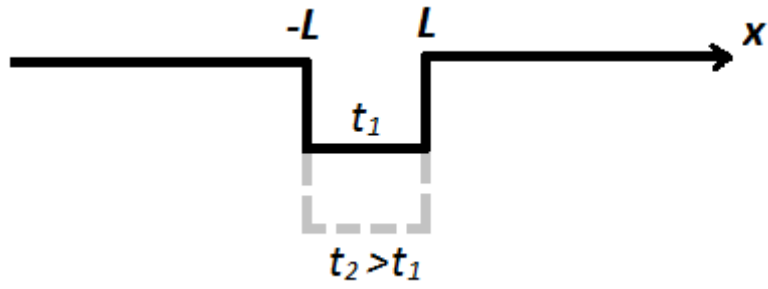

**Fig.2.** Scheme of the rectangular potential well deepening over time $t$.

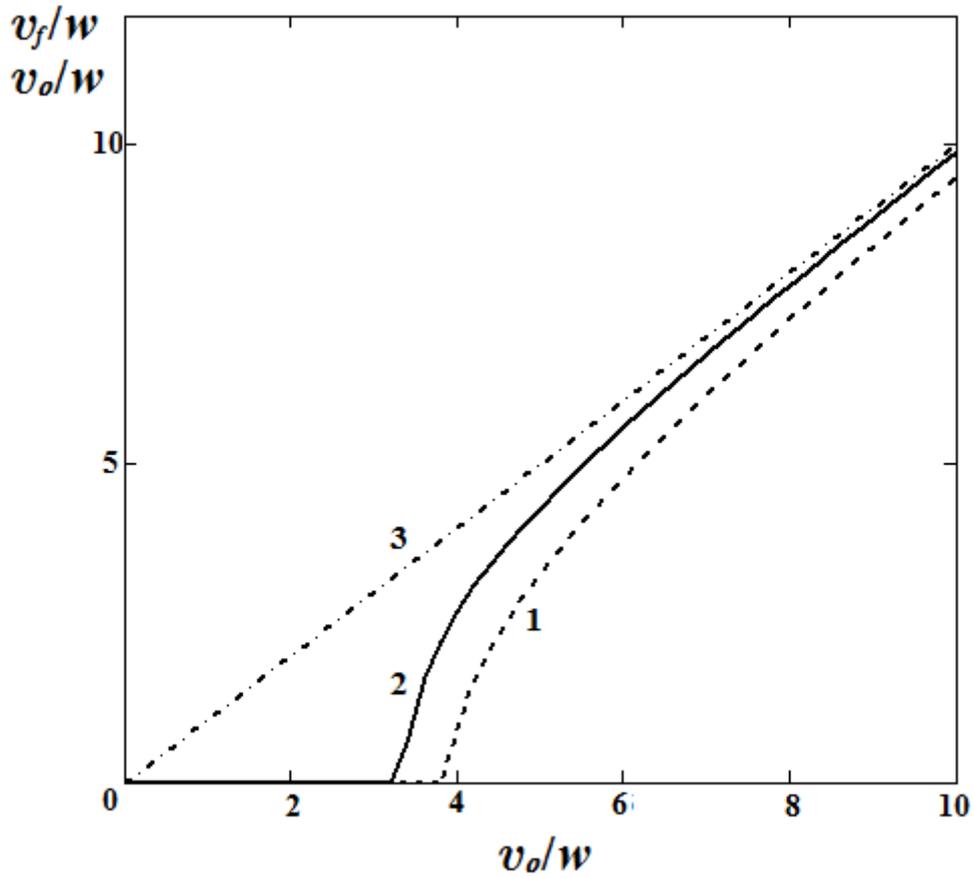

**Fig.3.** Dependence of the final speed $v_f$ (curves 1 and 2) of a particle, passing through the rectangular potential well (Fig.2), and also this particle initial velocity $v_0$ (straight line 3) on the value $v_0$ for the time moment $t=0.1T$ at the value $J_0 = 81K$ and parameter $n = 0.5$ (curve 1) and 2 (curve 2) of the function $\varphi(t)$ (13).

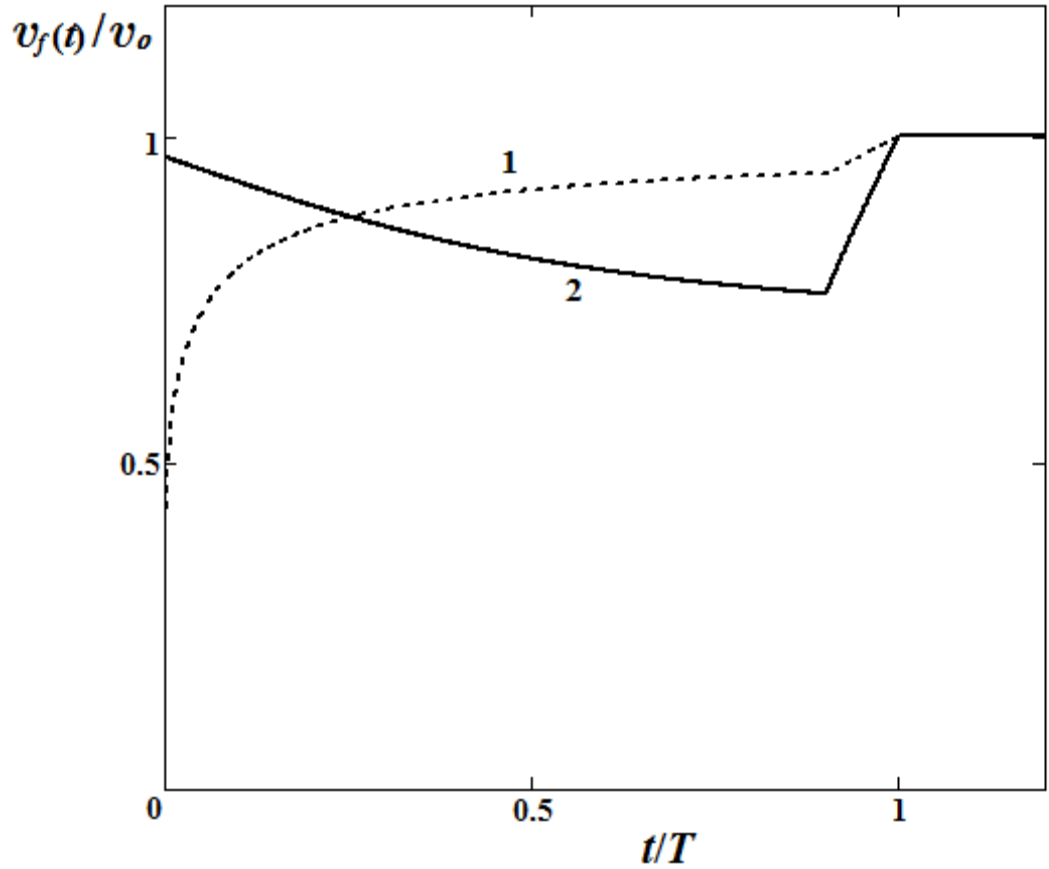

**Fig.4.** Dependence of the final speed $v_f(t)$ of particles, passing through the rectangular potential well (Fig.2), on the time moment $t$ of their getting into this well with the initial velocity $v_0 = 6\,w$ for the value $J_0 = 81K$ and parameter $n = 0.5$ (curve 1) and 2 (curve 2) of the function $\varphi(t)$ (13).